\documentclass[runningheads]{llncs}
\usepackage[T1]{fontenc}
\usepackage{cite}
\usepackage{wrapfig}
\usepackage{amsmath,amssymb,amsfonts}
\usepackage{algorithmic}
\usepackage{graphicx}
\usepackage{multirow}
\usepackage{nicematrix}
\usepackage{booktabs} 
\usepackage[normalem]{ulem} 
\usepackage{lipsum} 
\usepackage{subcaption}

\begin{document}
\title{Comparing Contrastive and Triplet Loss: Variance Analysis and Optimization Behavior}
\author{Donghuo Zeng\orcidID{0000-0002-6425-6270}}
\institute{KDDI Research, Inc., Japan \\
\email{do-zeng@kddi-research.jp}}
%

\maketitle              
\vspace{-10pt}
\begin{abstract}
Contrastive loss and triplet loss are widely used objectives in deep metric learning, yet their effects on representation quality remain insufficiently understood. We present a theoretical and empirical comparison of these losses, focusing on intra- and inter-class variance and optimization behavior (e.g., greedy updates). Through task-specific experiments with consistent settings on synthetic data and real datasets—MNIST, CIFAR-10—it is shown that triplet loss preserves greater variance within and across classes, supporting finer-grained distinctions in the learned representations. In contrast, contrastive loss tends to compact intra-class embeddings, which may obscure subtle semantic differences. To better understand their optimization dynamics, 
By examining loss‑decay rate, active ratio, and gradient norm, we find that contrastive loss drives many small updates early on, while triplet loss produces fewer but stronger updates that sustain learning on hard examples.
Finally, across both classification and retrieval tasks on MNIST, CIFAR-10, CUB-200, and CARS196 datasets, our results consistently show that triplet loss yields superior performance,  which suggests using triplet loss for detail retention and hard-sample focus, and contrastive loss for smoother, broad-based embedding refinement.
\end{abstract}

\section{Introduction}
Deep metric learning seeks to embed inputs into a space where geometric proximity reflects semantic similarity, enabling tasks such as image classification~\cite{Manmatha2017SamplingMI,schroff2015facenet} and image retrieval~\cite{Manmatha2017SamplingMI,li2015feature}. Two of the most popular margin‐based objectives are contrastive loss~\cite{hadsell2006dimensionality} and triplet loss~\cite{schroff2015facenet}. While both aim to maximize inter‐class separation, their different formulations yield distinct gradient patterns—and hence different “greediness” during training—that strongly influence embedding geometry and convergence~\cite{ghojogh2020fisher,musgrave2020metric}.

\paragraph{Why study gradient behavior?}  
Understanding how each loss allocates gradient effort—whether via many small, diffuse updates or fewer large, targeted steps—is crucial for tasks requiring fine‐grained retrieval or robust classification. A “greedy” loss will continue to enforce margins on easy samples, potentially over‐compacting clusters, whereas a more restrained update pattern may better preserve intra‐class diversity.

\paragraph{Variance structure}.  
We quantify how each loss is managed
1. \textit{Intra‐class variance}: dispersion of samples within a class, and  
2. \textit{Inter‐class variance}: separation margins between classes.  
Using overall and per‐class variance statistics, plus PCA projections of the original data vs.\ embeddings from each loss, we show that triplet loss maintains higher within‐class spread and clearer between‐class gaps on Synthetic data, MNIST, and CIFAR‑10, whereas contrastive loss tends to collapse clusters and blur subtle distinctions.
\vspace{-10pt}

\paragraph{Optimization greediness}  
We define \textit{greediness} via three metrics: \textit{Loss‐decay rate}: Epochs required to reduce loss by 90\% from the initial value, \textit{Active‐sample ratio}: fraction of pairs/triplets with nonzero gradient, \textit{Gradient norm}: average magnitude of parameter updates.  

On MNIST and CIFAR-10, contrastive loss reaches 90\% loss reduction by epoch 27, with a 65\% active-sample ratio and an average gradient norm of approximately 0.12—resulting in many small, diffuse updates and early convergence. In contrast, triplet loss requires until epoch 43 to reach the same reduction, with only 38\% active triplets but significantly larger gradient norms ($\approx$0.27), enabling more focused updates on hard examples and better preservation of embedding diversity.

Finally, we validate both losses on classification and retrieval tasks across MNIST, CIFAR‑10, CUB‑200, and CARS196, consistently finding that triplet loss outperforms contrastive loss. By formalizing variance analysis and greediness metrics, our study clarifies how each objective sculpts embedding geometry and training dynamics, and offers guidance on loss selection: use triplet loss for detail retention and hard‐sample emphasis, and contrastive loss for smoother, broad‐based refinement.

\section{Foundations of Contrastive and Triplet Loss}
\subsection{Contrastive and Triplet Loss}
Contrastive and triplet losses form the foundation of deep metric learning, where a neural network $\mathbf{f}(\cdot)$ maps inputs into an embedding space so that semantically similar samples are close and dissimilar ones are separated. We simplify notation by using a single embedding function $\mathbf{f}$ for all inputs.

\textit{Contrastive Loss:} Originally proposed by Hadsell \emph{et al.}~\cite{hadsell2006dimensionality}, contrastive loss is
\begin{align}
\mathcal{L}_{\text{con}}
= \sum_{(x,y)\in P} \|\mathbf{f}(x)-\mathbf{f}(y)\|^2 
\;+\; \sum_{(x,y)\in N} \bigl[\,m - \|\mathbf{f}(x)-\mathbf{f}(y)\|\,\bigr]_+^2,
\end{align}
where $P$ and $N$ are sets of positive and negative pairs, $m>0$ is the margin enforcing a minimum separation that controls the trade-off between intra-class compactness and inter-class separation. A larger $m$ encourages greater inter-class distances but may permit more intra-class variance, while a smaller $m$ enforces tighter clusters. $[z]_+=\max(0,z)$. $\|\cdot\|$ denotes the L2 norm. By independently pulling every positive pair together and pushing every negative pair apart—even after the margin is met—contrastive loss exhibits a “greedy” optimization behavior, resulting in many small gradient updates across samples~\cite{musgrave2020metric}.

\textit{Triplet Loss:} Introduced in FaceNet by Schroff \emph{et al.}~\cite{schroff2015facenet}, triplet loss uses triplets $(a,p,n)$ of anchor, positive, and negative:
\begin{equation}
\centering
\begin{aligned}
\mathcal{L}_{\text{tri}}
= \sum_{(a,p,n)} \bigl[\,\|\mathbf{f}(a)-\mathbf{f}(p)\|^2 - \|\mathbf{f}(a)-\mathbf{f}(n)\|^2 + m\,\bigr]_+.
\end{aligned}
\end{equation}
Here, the same margin $m$ ensures the anchor–positive distance is at least $m$ smaller than the anchor–negative distance. Once a triplet satisfies this ranking constraint, it no longer contributes gradients, yielding fewer but larger updates focused on hard examples~\cite{hermans2017in, wang2019multi}. The key conceptual differences include: 
(1) \emph{Intra-class dispersion control}: Contrastive loss can collapse within-class samples under a fixed margin; triplet loss permits richer spread~\cite{musgrave2020metric}. 
(2) \emph{Inter-class margin enforcement}: Contrastive loss enforces a hard absolute gap; triplet loss ensures only relative separation. 
(3) \emph{Greedy optimization behavior}: Contrastive loss continues to update all pairs post-margin, resulting in frequent low-magnitude updates. Triplet loss applies gradients only to violating triplets, concentrating updates on harder examples (Section~\ref{sec:greediness_in_optimization}).
(4) \emph{Ranking vs.\ Absolute Distance}: Triplet loss’s ranking formulation makes it particularly suited to retrieval tasks (e.g., face or product retrieval~\cite{hermans2017in}), where preserving relative similarities is paramount.

\subsection{Variance and Optimization Greediness}
Maintaining an appropriate structure in the embedding space—where samples of the same class are compact yet not collapsed, and different classes remain well-separated—is essential for both fine-grained retrieval and classification. To quantify this structure, we compute  intra-class and inter-class variances as follows:
\begin{equation}
\centering
\begin{aligned}
\sigma_{\mathrm{intra}}^2 &= \frac{1}{C} \sum_{c=1}^C \frac{1}{N_c} \sum_{i \in I_c} \| z_i - \mu_c \|^2, \quad \text{where} \quad \mu_c = \frac{1}{N_c} \sum_{i \in I_c} z_i, \quad z_i = \mathbf{f}(x_i), \\
\sigma_{\mathrm{inter}}^2 &= \frac{1}{C(C-1)} \sum_{c \neq c'} \| \mu_c - \mu_{c'} \|^2,
\end{aligned}
\end{equation}

where $C$ is the number of classes, $I_c$ is the set of indices for class $c$, and $N_c = |I_c|$. Here, $\sigma_{\mathrm{intra}}^2$ measures the average spread of embeddings within each class, while $\sigma_{\mathrm{inter}}^2$ quantifies the average separation between the centroids of the classes.

\paragraph{Optimization greediness} 
We term \emph{greediness} the propensity of a loss to continue optimizing already-satisfied constraints, potentially leading to excessive intra-class compaction—where embeddings within a class become overly concentrated—or even dimensional collapse, where the embedding space reduces to a lower-dimensional subspace~\cite{Jing2021UnderstandingDC}. 
When measured by 
(1) $\textit{Loss-decay rate} = \min \left\{\, e \;\middle|\; \mathcal{L}^{(e)} \leq 0.1 \cdot \mathcal{L}^{(0)} \,\right\},] $, where \(\mathcal{L}^{(e)}\) denotes the average loss at epoch \(e\). This measures how quickly the loss decreases to 10\% of its initial value and captures coarse convergence speed. 
(2) \(\textit{Active Ratio} = \frac{|\{(x,y) \in P \cup N : \mathcal{L}_{\text{con/tri}}(x,y) > 0\}|}{|\text{Batch}|}\), the fraction of samples with nonzero loss per batch~\cite{Manmatha2017SamplingMI}. This quantifies how many samples continue to drive learning.  
(3) \emph{gradient norm} computed as the $\mathcal{L}2$ norm of the loss gradient: (\(\|\nabla \mathcal{L}\|_2\)), measuring the overall magnitude of parameter updates.

Contrastive loss typically shows a high active ratio and low gradient norm, leading to widespread low-magnitude updates across the batch—even when constraints are already satisfied~\cite{musgrave2020metric, Manmatha2017SamplingMI}. Triplet loss tends to activate fewer samples but produces stronger gradients concentrated on difficult examples~\cite{hermans2017in, Manmatha2017SamplingMI}. These distinct behaviors reflect deeper trade-offs between convergence speed and structural preservation, explored further in Section~\ref{sec:greediness_in_optimization}.

\section{Experimental Framework}
\subsection{Datasets}
\subsubsection{Synthetic data}
We generate synthetic data in a fixed 128-dimensional space with 10 clusters, each containing 200 samples, plus Gaussian outliers. 

\begin{enumerate}
\item \textbf{Class centers:}  
    For each class \(c\in\{1,\dots,10\}\), draw
    \[
      g_c \sim \mathcal{N}(0,I_{128}),
      \qquad
      \mu_c = 5\,g_c
    \]
    so that each center has covariance \(25I_{128}\).
  \item \textbf{Covariance and noise:}  
    For each class \(c\), sample a random matrix \(A_c\sim\mathcal{N}(0,I_d)\) and set
    \(\Sigma_c = A_cA_c^\top\). Compute the Cholesky factor \(L_c\) of \(\Sigma_c + 10^{-3}I_d\).  
    Then for each of the 200 points:
    \[
      z_i \sim \mathcal{N}(0,I_{128}),
      \quad
      n_i = L_c\,z_i,
      \quad
      x_i = \mu_c + 1.4\,n_i.
    \]
 \item \textbf{Label overlap (probability \(p=0.1\)):}  
    Assign each point label \(c\), but with probability 0.1 reassign it to a random class in \(\{1,\dots,10\}\).
 \item \textbf{Gaussian outliers (fraction 0.05):}  
    After sampling all cluster points,
    \[
      n_{\text{outliers}} = \left\lfloor 10 \times 200 \times 0.05 \right\rfloor = 100
    \]
    points are appended that drawn from \(\mathcal{N}(0,15^2I_{128})\), each labeled as \(-1\).
\end{enumerate}
\vspace{-10pt}
\subsubsection{Real dataset}
We evaluate on two classification and three retrieval datasets, with CIFAR-10, CARS196, and CUB-200 embeddings extracted via Vision Transformer (ViT)\footnote{https://huggingface.co/docs/transformers/v4.13.0/en/model\_doc/vit}. (1) \emph{MNIST:} 10 classes, grayscale 28\(\times\)28 images; 60,000 training and 10,000 test samples (\(\sim6,000\) per class in training, \(\sim1,000\) per class in testing). (2) \emph{CIFAR-10:}  10 classes, RGB 32\(\times\)32 images; 50,000 training and 10,000 test samples (5,000 per class in training, 1,000 per class in testing). (3) \emph{CARS196:}  196 classes fine-grained categories; 8,144 training and 8,041 test images (\(\sim42\) images per class). (4) \emph{CUB-200:}  200 fine-grained bird categories; 11,788 images (5,994 for training, 5,794 for testing).

\subsection{Training Details}
\emph{Model Architectures:}
Synthetic data leverages a simple MLP: two fully connected layers (128$\rightarrow$64$\rightarrow$32) with ReLU activations and L2 normalization on the 32-D output. Real data (MNIST, CIFAR-10) uses a CNN-like model: two Conv-ReLU-MaxPool blocks followed by linear layers (flatten$\rightarrow$128$\rightarrow$64) and L2-normalized embeddings. Retrieval tasks with CIFAR-10, CARS196, and CUB-200 use a frozen ViT-B/32 backbone with a 512-D embedding head and L2 normalization.
\emph{Optimization Setup:}
All models are trained with Adam (learning rate=1e-3, weight decay=1e-5), batch size=64, for 50 epochs, and margin \(m=1.0\) for both losses. Euclidean distance is used for all pairwise comparisons.
\emph{Loss Sampling Strategies:}
For both contrastive and triplet loss, we sample 50\% positive and 50\% negative pairs, excluding outliers (label \(-1\)) from positives.
\emph{Diagnostics and Visualization:}
We track loss curves, active ratio (fraction of non-zero losses per batch), and gradient norms to analyze optimization dynamics. Embedding visualization is performed via PCA. Code availability \footnote{https://anonymous.4open.science/r/tc-2025}

\section{Results and Analysis}
\subsection{Variances analysis}
\vspace{-30pt}
\begin{table}[h]
\centering
\caption{statistics of intra- and inter-class variance  on Synthetic data and MNIST}
\begin{tabular}{lcccccccc}
\toprule
\multirow{2}{*}{Loss} & \multicolumn{4}{c}{Synthetic data} & \multicolumn{4}{c}{MNIST} \\ 
\cmidrule(lr){2-5} \cmidrule(lr){6-9}
& \multicolumn{2}{c}{Intra-class} & \multicolumn{2}{c}{Inter-class} & \multicolumn{2}{c}{Intra-class} & \multicolumn{2}{c}{Inter-class} \\ 
\cmidrule(lr){2-3} \cmidrule(lr){4-5} \cmidrule(lr){6-7} \cmidrule(lr){8-9}
& $\mu$ & $\sigma^2$ & $\mu$ & $\sigma^2$ & $\mu$ & $\sigma^2$ & $\mu$ & $\sigma^2$ \\
\midrule
Contrastive   & 0.031   & 6.4e-05 & 1.2149 & 0.0710& 0.0030 & 0.0001 & 1.0347 & 0.0064 \\
Triplet       & 0.074  & 0.0001 & 1.4399 & 0.0342 & 0.0059 & 0.0001 & 1.4840 & 0.0047 \\
\bottomrule
\end{tabular}\\
\footnotesize{* paired t--test, $p < 0.001$}\\
\label{tab:variance_con_tri}
\vspace{-20pt}
\end{table}

\begin{wrapfigure}{r}{0.55\linewidth}
\vspace{-20pt}
  \centering
  \includegraphics[width=\linewidth]{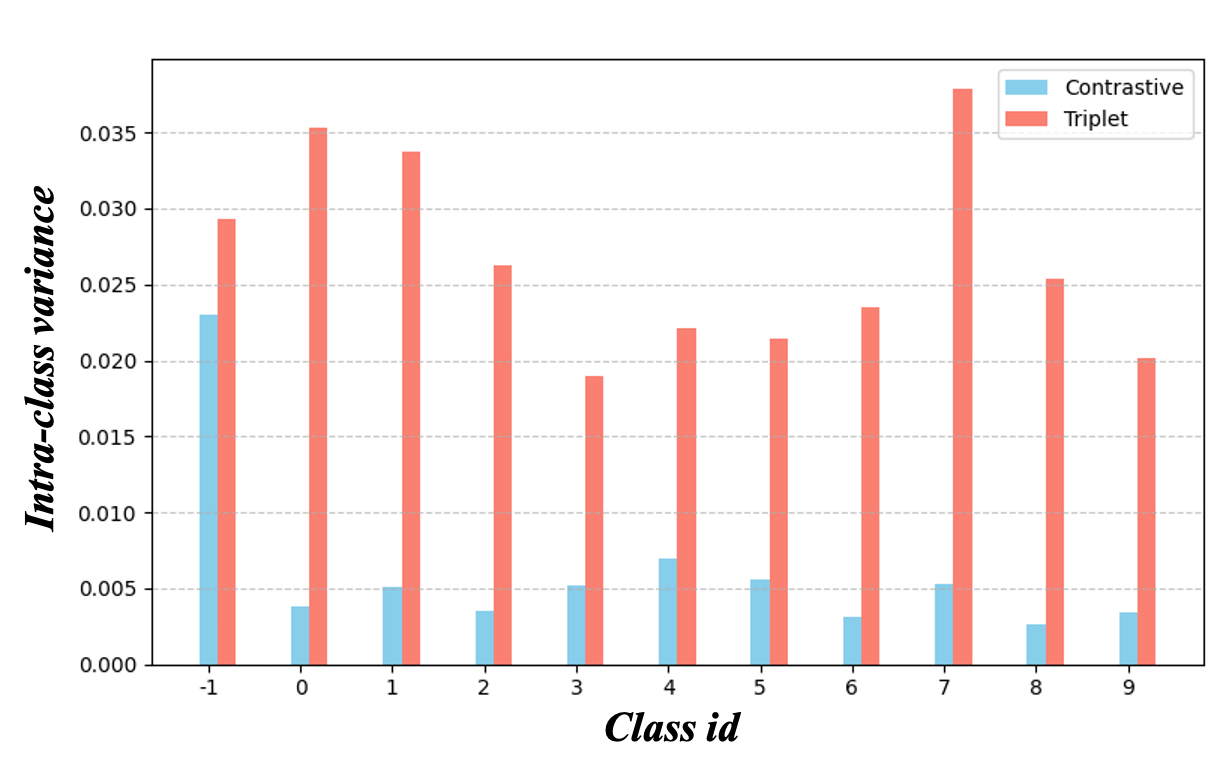}
  \caption{Intra-class variance for each class}
  \label{fig:intra_class_var_each}
  \vspace{-25pt}
\end{wrapfigure}
\begin{figure}[t]
\vspace{-20pt}
  \centering
  \includegraphics[width=0.9\linewidth]{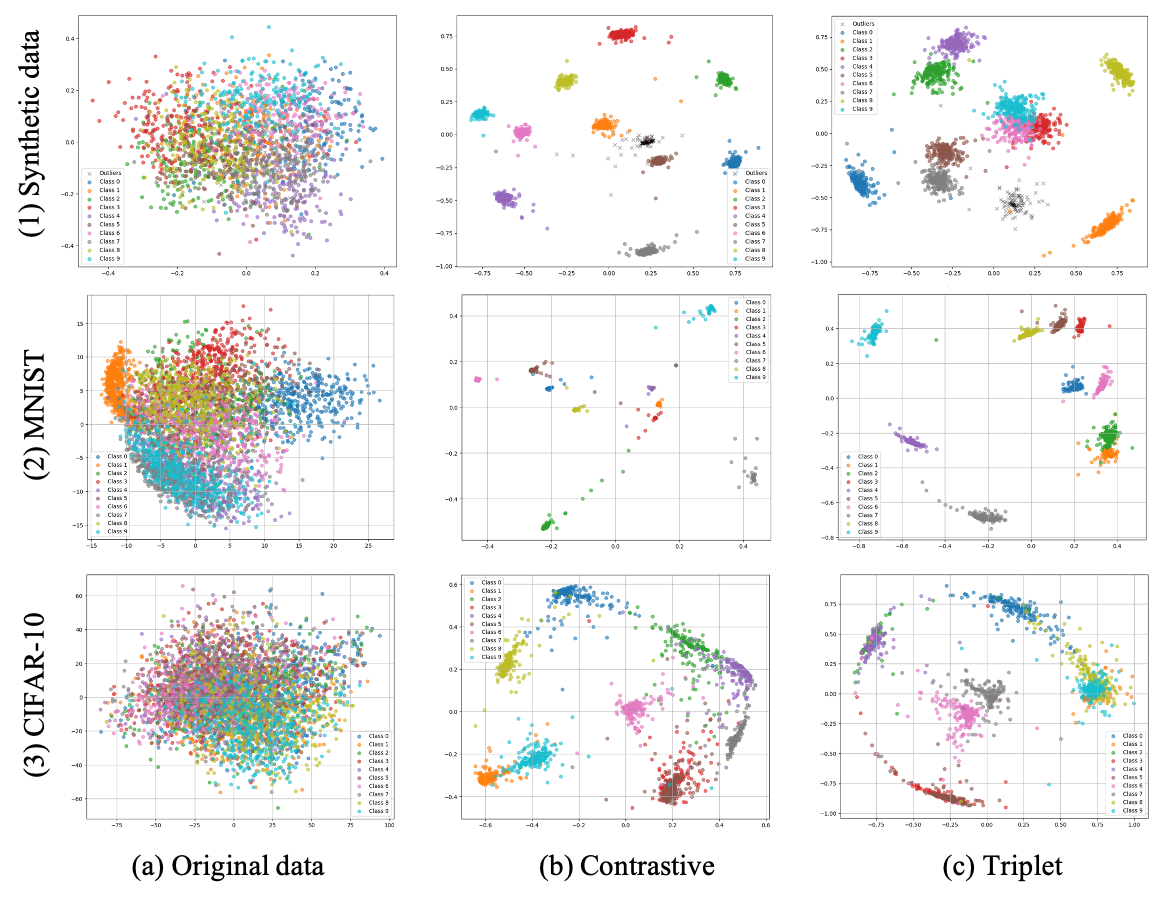}
  \caption{PCA embedding visualizations of three distinct datasets}
  \label{fig:data_contrastive_triplet}
  \vspace{-15pt}
\end{figure}

We can observe in Table~\ref{tab:variance_con_tri} that on Synthetic data, the triplet loss preserves approximately 2.4 times more average intra-class variance than the contrastive loss (0.074/0.031 $\approx$2.4). This indicates that embeddings trained with triplet loss exhibit greater within-class diversity. A paired $t$-test on per-class intra-class variances confirms this difference is statistically significant ($p < 0.001$). 
The average inter-class distance is slightly higher for triplet loss (1.4399 vs. 1.2149), with lower variability in inter-class distances ($\sigma^{2}$=0.0342 vs. 0.0710), suggesting more consistent separation between classes. As illustrated in Fig.~\ref{fig:intra_class_var_each}, the intra-class variances under triplet loss are uniformly higher across all classes compared to contrastive loss.
A similar trend is observed for the MNIST dataset. These traits—greater within-class diversity and clearer class separation—make triplet loss well-suited for downstream tasks needing robust embedding generalization~\cite{schroff2015facenet, musgrave2020metric}. Fig.~\ref{fig:data_contrastive_triplet} shows PCA projections of the learned embeddings. In general, \textit{contrastive loss} yields tightly clustered classes, while \textit{triplet loss} allows for more natural, dispersed clusters that better reflect the underlying data structure.

\subsection{Greedy Optimization Behavior of Loss Functions}
\label{sec:greediness_in_optimization}
\begin{figure}
    \centering
    \begin{subfigure}{0.32\linewidth}
    \includegraphics[width=\linewidth]{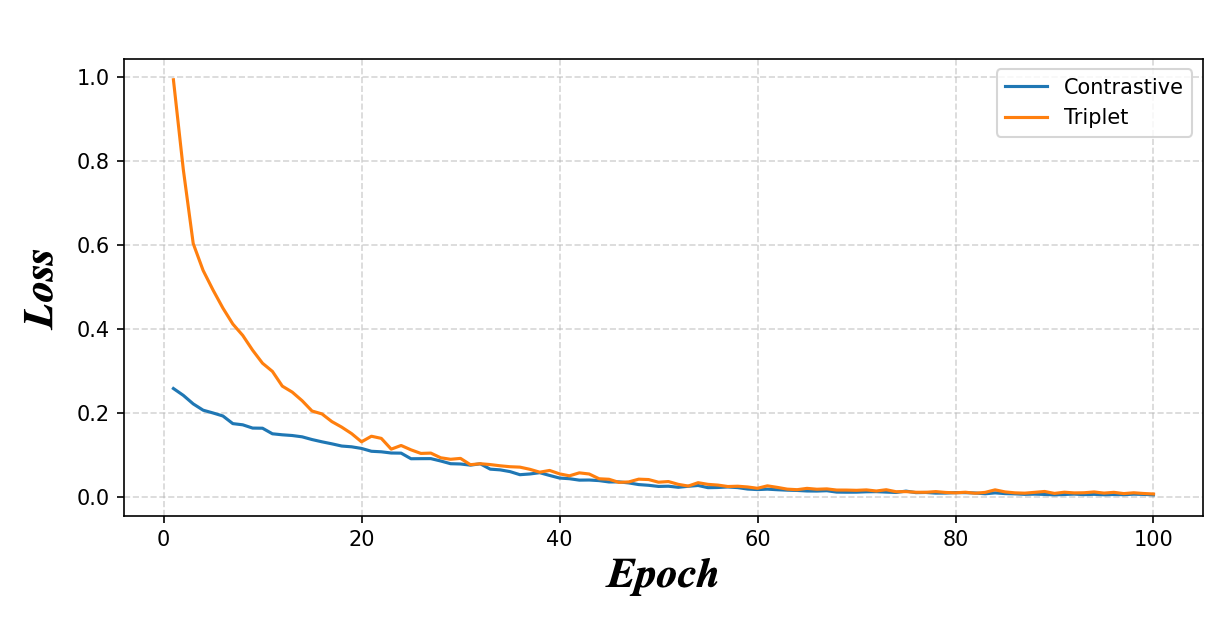}
    \caption{Loss Value}
    \label{fig:loss_value}
    \end{subfigure}
    \begin{subfigure}{0.32\linewidth}
    \centering
    \includegraphics[width=\linewidth]{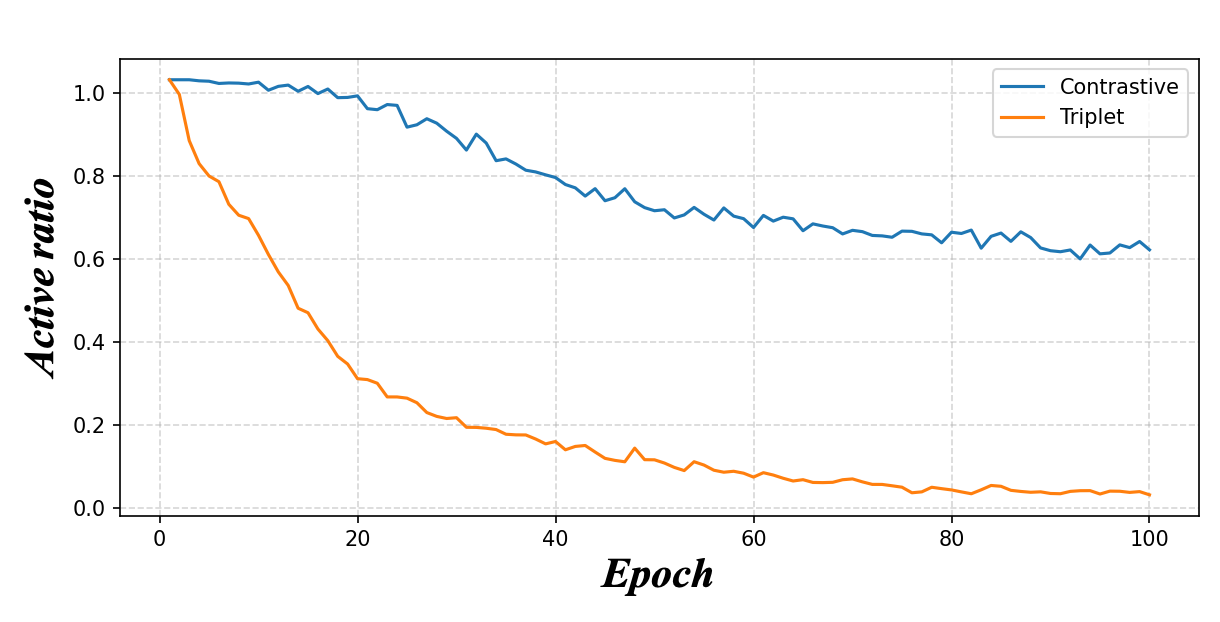}
    \caption{Active Ratio}
    \label{fig:intra_class_var1}
    \end{subfigure}
    \begin{subfigure}{0.32\linewidth}
    \centering
    \includegraphics[width=\linewidth]{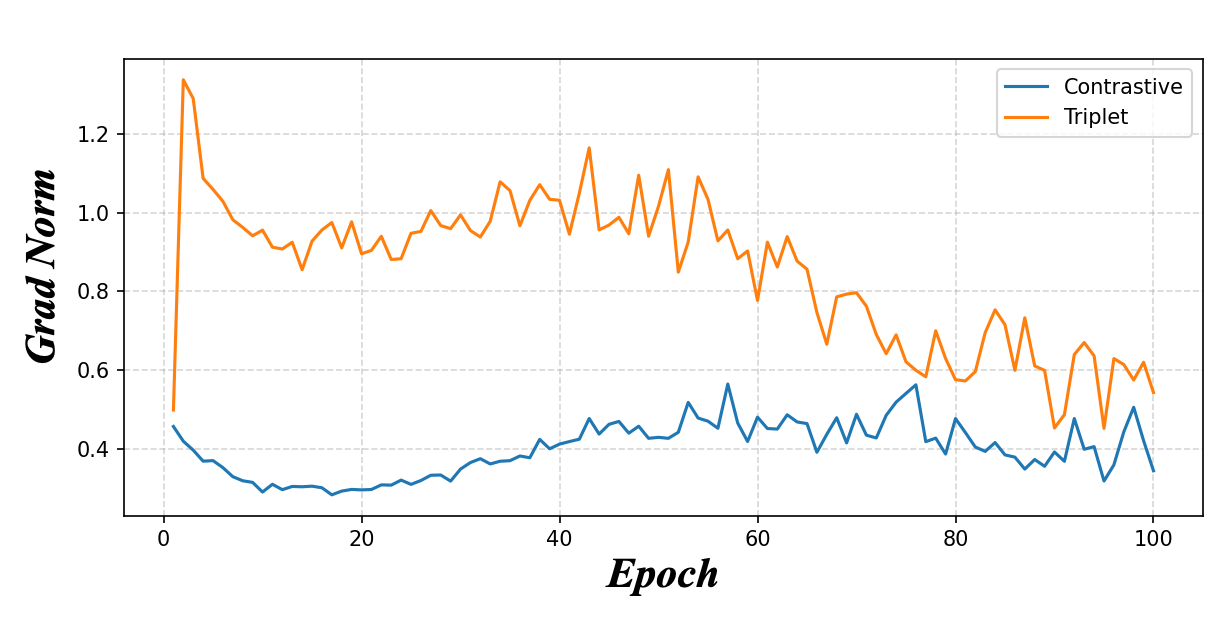}
    \caption{Gradient Norm}
    \label{fig:intra_class_var2}
    \end{subfigure}
    \caption{The greediness metrics over training epochs.}
    \label{fig:greediness_metrics}
    \vspace{-20pt}
\end{figure}

\begin{wraptable}{r}{0.5\textwidth}
\vspace{-20pt}
  \centering
  \setlength{\tabcolsep}{3pt} 
  \caption{Greediness metrics at epoch 100}
  \vspace{-10pt}
  \small
  \begin{tabular}{lcc}
    \toprule
    Metric & Contrastive & Triplet \\
    \midrule
    Active ratio & 65\% & 38\% \\
    Gradient norm & 0.12 & 0.27 \\
    Loss-decay rate & 27 & 43 \\
    \bottomrule
  \end{tabular}
  \label{tab:greedy_metrics}
  \vspace{-20pt}
\end{wraptable}
Different metric‐learning losses induce distinct update patterns—what we term “greediness”—measured by three metrics: Loss-decay rate from loss curve , active ratio, and gradient norm. Table~\ref{tab:greedy_metrics}, we briefly report Loss-decay rate (the epoch by which 90\% of the initial loss is eliminated) and focus on comparing results: Contrastive loss reaches 90\% loss reduction by epoch 27, engages a large share of samples (65\% active ratio), and shows modest gradients (norm $\approx$ 0.12). This combination yields many small, diffuse updates and early plateauing of training. Triplet loss, however, achieves 90\% Loss-decay only by epoch 43, with fewer active triplets (38\%) but stronger updates (norm $\approx$ 0.27). These sharper, focused updates prolong learning and help preserve fine‐grained distinctions in the embedding space. Fig.~\ref{fig:greediness_metrics} illustrates that \emph{contrastive loss} causes a sharper and earlier collapse in intra-class distances, converging faster to a low-variance state than \emph{triplet loss}. This delayed decay in triplet loss, however, sustains learning on harder examples, which helps explain its superior early retrieval performance (see Section~\ref{sec:classification_retrieval}). 

\subsection{Application to Classification and Retrieval}
\label{sec:classification_retrieval}
\begin{wraptable}{r}{0.45\textwidth} 
\vspace{-20pt}
\centering
\caption{Classification accuracy on MNIST and CIFAR-10}
\begin{tabular}{lcc}
\toprule
Loss & MNIST & CIFAR-10 \\
\midrule
Contrastive   & 0.9869 & 0.8998 \\
Triplet       & 0.9933 & 0.9371 \\
\bottomrule
\end{tabular}
\label{tab:classification}
\vspace{-20pt}
\end{wraptable}
To demonstrate the practical impact of our variance‐structure and greedy‐optimization analyses, we evaluate both classification and retrieval tasks. Classification tests global separation—requiring clear inter‐class margins and compact clusters—while retrieval measures fine‐grained neighbor ranking—benefiting from preserved intra‐class variance. Together, they link embedding geometry and optimization dynamics to real‐world performance.

Table~\ref{tab:classification} shows that triplet loss achieves higher classification accuracy (MNIST: 0.9933 vs.\ 0.9869; CIFAR-10: 0.9371 vs.\ 0.8998), while Table~\ref{tab:retrieval} demonstrates its superior retrieval r@1 performance across CIFAR-10 (0.9192 vs.\ 0.8433), CARS196 (0.2982 vs.\ 0.2542), and CUB-200 (0.3421 vs.\ 0.3154), with smaller gaps at r@5 and r@10. These results confirm that triplet’s broader intra-class variance preserves fine distinctions—boosting r@1—while still enforcing inter-class margins for high accuracy. In contrast, contrastive’s many small, rapid updates over-compact clusters and hurt both retrieval and separability. Balancing intra-class spread with update intensity is therefore key to optimal classification and retrieval.

\begin{table}[h]
\vspace{-20pt}
\centering
\caption{Retrieval recall@k (k=1,5,10) on three datasets.}
\begin{tabular}{lccccccccc}
\toprule
\multirow{2}{*}{Loss} & \multicolumn{3}{c}{CIFAR-10} & \multicolumn{3}{c}{CARS196} &\multicolumn{3}{c}{CUB-200}\\ 
\cmidrule(lr){2-4} \cmidrule(lr){5-7} \cmidrule(lr){8-10}
& r@1 & r@5 & r@10 & r@1 & r@5 & r@10 & r@1 & r@5 & r@10\\
\midrule
Contrastive &0.8433 &0.9701 &0.9899
 &  0.2542 &  0.5249 &  0.6596 
 &0.3154 &0.5489 &0.6897  \\
Triplet     & 0.9192 &0.9694 &0.9793 & 0.2982 &0.5540 &0.6667 &  0.3421 &  0.5876 &  0.7234
\\ 
\bottomrule
\end{tabular}
\label{tab:retrieval}
\vspace{-10pt}
\end{table}

\section{Conclusion}
We presented a theoretical and empirical comparison of contrastive and triplet loss in deep metric learning, focusing on both embedding structure and optimization behavior. Our variance analysis shows that triplet loss preserves greater intra- and inter-class variance, supporting finer-grained distinctions, while contrastive loss tends to compact intra-class representations. Through metrics such as loss-decay rate, active ratio, and gradient norm, we also find that contrastive loss applies frequent small updates, whereas triplet loss produces fewer but stronger updates, concentrating learning on hard examples. Across classification and retrieval tasks, triplet loss consistently outperforms contrastive loss. These findings suggest that triplet loss is better suited for detail-preserving, discriminative embeddings, while contrastive loss favors smoother, broad-based representation learning. Future work includes exploring hybrid losses and adaptive margins that better balance precision and generalization.

\bibliography{refs}

\begin{thebibliography}{1}

\bibitem{ghojogh2020fisher}
Benyamin Ghojogh, Milad Sikaroudi, Sobhan Shafiei, Hamid~R Tizhoosh, Fakhri Karray, and Mark Crowley.
\newblock Fisher discriminant triplet and contrastive losses for training siamese networks.
\newblock In {\em 2020 international joint conference on neural networks (IJCNN)}, pages 1--7. IEEE, 2020.

\bibitem{hadsell2006dimensionality}
Raia Hadsell, Sumit Chopra, and Yann LeCun.
\newblock Dimensionality reduction by learning an invariant mapping.
\newblock In {\em 2006 IEEE Computer Society Conference on Computer Vision and Pattern Recognition (CVPR'06)}, pages 1735--1742. IEEE, 2006.

\bibitem{hermans2017in}
Alexander Hermans, Lucas Beyer, and Bastian Leibe.
\newblock In defense of the triplet loss for person re-identification.
\newblock In {\em Proceedings of the IEEE International Conference on Computer Vision (ICCV)}, pages 390--398, 2017.

\bibitem{Jing2021UnderstandingDC}
Li~Jing, Pascal Vincent, Yann LeCun, and Yuandong Tian.
\newblock Understanding dimensional collapse in contrastive self-supervised learning.
\newblock {\em ArXiv}, abs/2110.09348, 2021.

\bibitem{li2015feature}
Wu-Jun Li, Sheng Wang, and Wang-Cheng Kang.
\newblock Feature learning based deep supervised hashing with pairwise labels.
\newblock {\em arXiv preprint arXiv:1511.03855}, 2015.

\bibitem{Manmatha2017SamplingMI}
R.~Manmatha, Chaoxia Wu, Alex Smola, and Philipp Kr{\"a}henb{\"u}hl.
\newblock Sampling matters in deep embedding learning.
\newblock {\em 2017 IEEE International Conference on Computer Vision (ICCV)}, pages 2859--2867, 2017.

\bibitem{musgrave2020metric}
Kevin Musgrave, Serge Belongie, and Ser-Nam Lim.
\newblock A metric learning reality check.
\newblock In {\em Computer Vision--ECCV 2020: 16th European Conference, Glasgow, UK, August 23--28, 2020, Proceedings, Part XXV 16}, pages 681--699. Springer, 2020.

\bibitem{schroff2015facenet}
Florian Schroff, Dmitry Kalenichenko, and James Philbin.
\newblock Facenet: A unified embedding for face recognition and clustering.
\newblock In {\em Proceedings of the IEEE conference on computer vision and pattern recognition}, pages 815--823, 2015.

\bibitem{wang2019multi}
Qian Wang, Jian Wang, Wen Liu, Yichao Gao, and Yongdong Xu.
\newblock Multi-similarity loss with general pair weighting for deep metric learning.
\newblock In {\em CVPR}, pages 5022--5030, 2019.

\end{thebibliography}
\bibliographystyle{plain}
\end{document}